\documentclass[%
 reprint,
 amsmath,amssymb,
 aps,
floatfix,
]{revtex4-2}

\usepackage{graphicx}
\usepackage{dcolumn}
\usepackage{bm}
\usepackage{subcaption}
\usepackage{xcolor}
\usepackage{braket}
\usepackage{amsmath,amsthm,amssymb}
\usepackage{hyperref}
\hypersetup{
    colorlinks=true,
    linkcolor=blue,
    urlcolor=red,
    linktoc=all
}

\usepackage{titlesec} 
\usepackage{mhchem}

\titleformat{\subsection}[runin]
  {\normalfont\large\bfseries}{\thesubsection}{1em}{}
\titleformat{\subsubsection}[runin]
  {\normalfont\normalsize\bfseries}{\thesubsubsection}{1em}{}
  
\newcommand{\ts}{\Tilde{\sigma}}
\newcommand{\tr}{\Tilde{r}}
\newcommand{\Erf}{\text{Erf}}

\newcommand{\comment}[1]{}

\begin{document}

\preprint{APS/123-QED}

\title{Analysis of Fr{\"o}hlich bipolarons}

\author{Lisa Lin}
\affiliation{%
 Department of Physics, University of Chicago, Chicago, Illinois 60637, USA
}%

\author{Peter B. Littlewood}%
\author{Alex Edelman}
\affiliation{%
James Franck Institute and Department of Physics, University of Chicago, Chicago, Illinois 60637, USA \\
Physical Sciences and Engineering, Argonne National Laboratory, Lemont, Illinois 60439, USA
}%

\date{\today}

\begin{abstract}
Following a resurgence of interest in dilute superconductivity in polar semiconductors, we perform a variational calculation to probe the existence of Fr{\"o}hlich bipolarons in these materials. Our solution is capable of interpolating between the weak- and strong-coupling limits of the electron-phonon interaction. We predict bipolaron formation in solely the strong-coupling regime, and we are not aware of any existing materials at these parameter values. However, imposing an extrinsic electron size constraint to mimic confinement on the sub-micron scale produces binding in new parts of the phase diagram, including at weak coupling, within reach of the near-ferroelectric perovskites.
\end{abstract}

\maketitle

\section{\label{sec:level1}Introduction}
An electron moving through a polar crystal displaces the positive and negative ions around it; these dynamical lattice distortions (phonons) generate a potential well around the electron and modify its dynamics. The electron dressed by its cloud of phonons is known as a polaron. Two polarons may form a bipolaron if the phonon-mediated attraction dominates over the Coulomb repulsion. We here focus on polarons and bipolarons which are large in the sense of Emin \cite{Emin89}, denoting quasiparticles formed by long-range Coulombic interactions at zero temperature, and ignore details on the scale of the lattice constant. In this context, longitudinal optic (LO) phonons coupling to charge carriers provide the dominant attractive forces involved. 

\subsubsection{Polarons}
There is a rich history of variational theories of the ``slow'' polaron, i.e. in the regime in which the electron is anti-adiabatic with respect to the phonons \cite{LLP,feyn55,Huy77}. These theories notably all have as a single governing parameter the dimensionless Fr{\"o}hlich coupling constant $\alpha$, and can be classified as ``weak-coupling'' or ``strong-coupling'' solutions for ground state energies $E_{GS} \propto \alpha$ and $E_{GS} \propto \alpha^2$, respectively. Feynman \cite{feyn55} and Huybrechts \cite{Huy77} obtain solutions that interpolate between these two limits. Although many variational theories find a discontinuous transition between weak- and strong-coupling regimes, it has been rigorously shown that only a crossover exists \cite{PeetersDevreesePhaseTransition,GerlachLowenPhaseTransition}, as also observed in diagrammatic Monte Carlo calculations \cite{Svistunov2000}. Ref.~\cite{LDB77} generalizes the transformation of Lee, Low, and Pines \cite{LLP} (LLP) to the context of an interacting polaron gas; we will apply the same transformation to our problem in the following sections. 

\subsubsection{Bipolarons}
Theories of bipolarons arise in many contexts, including bipolarons in Bose Einstein condensates (BEC) \cite{Camacho-Guardian18} and conducting polymers \cite{Bredas85}. Much recent interest in bipolarons has centered around bipolaronic superconductivity; this topic has been studied extensively in Ref.~\cite{EminHillery89}. (See Refs.~\cite{Alexandrov1994,Devreese_2009} for reviews on polarons, bipolarons, and bipolaronic superconductivity.) Bipolaron formation is controlled by a competition between Coulomb repulsion with dimensionless coupling strength $U$, and phonon-mediated attraction where the electron-phonon attraction is measured by the Fr{\"o}lich coupling constant $\alpha = U(1-\eta)/2$. $\alpha$ depends both on the bare electron-ion Coulomb coupling and the polarizability of the crystal, the latter of which is parameterized by the ratio of dielectric constants $\eta = \varepsilon_\infty/\varepsilon_0$.

Variational theories of bipolarons in particular draw upon elements of the polaron theories listed above. \cite{Devreese91} and \cite{devreese92} employ Feynman path integral methods. Refs.~\cite{devreese92,Devreese1994} use variational methods to perform strong-coupling analyses and predict bipolaron formation below certain critical values of $\eta$. Ref.~\cite{Lakhno10} performs an analogous calculation for a large-radius bipolaron. Ref.~\cite{Adamowski_1992} discusses the effect of self-trapping (i.e. localized wave functions) on the stability of Fr{\"o}lich bipolarons.

\begin{figure}[h!]
  \centering
  \begin{subfigure}{\linewidth}
    \includegraphics[width=\linewidth]{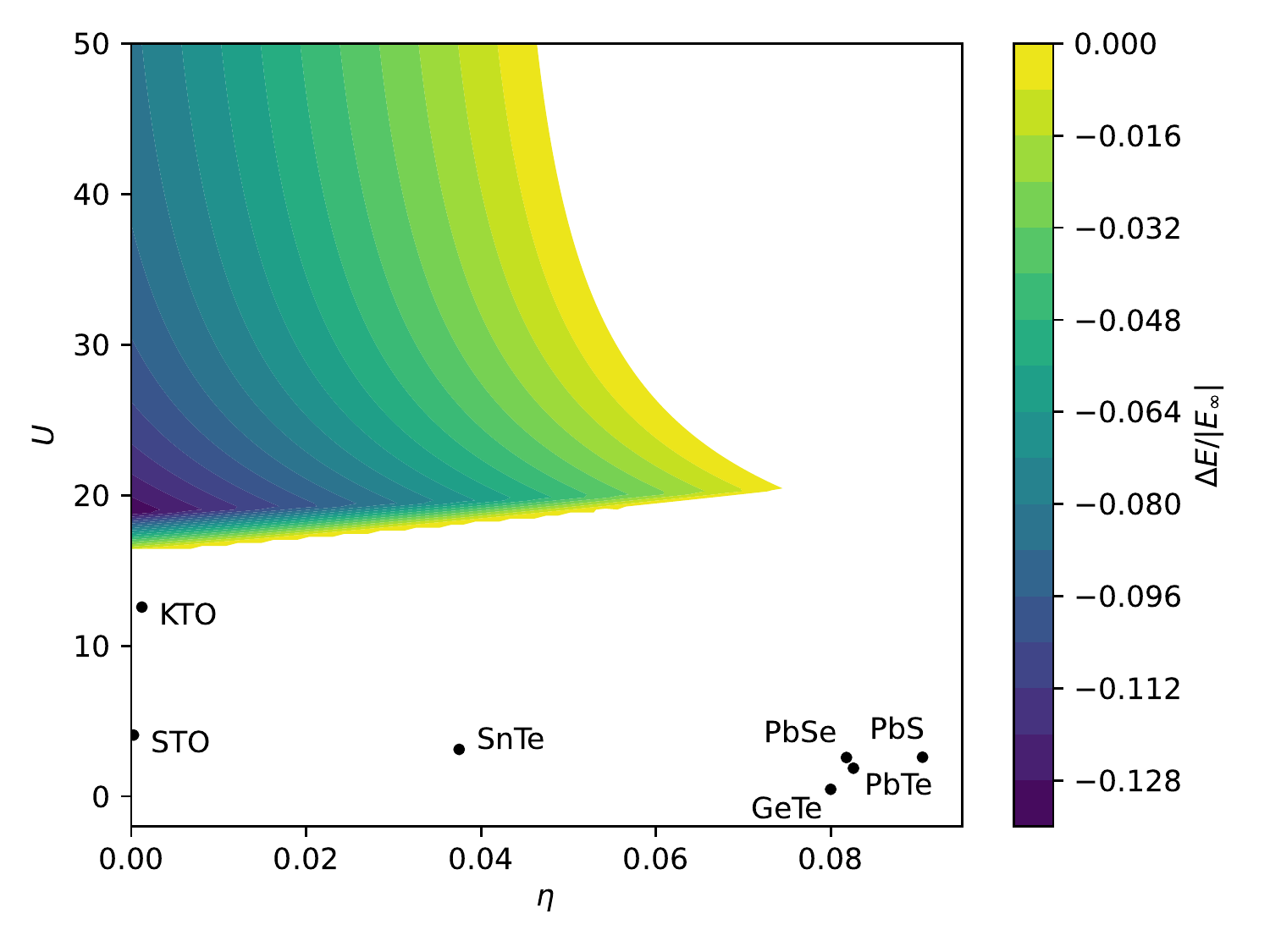}
    \caption{}
    \label{fig:nak_PD_full}
  \end{subfigure}\\
  \begin{subfigure}{\linewidth}
    \includegraphics[width=\linewidth]{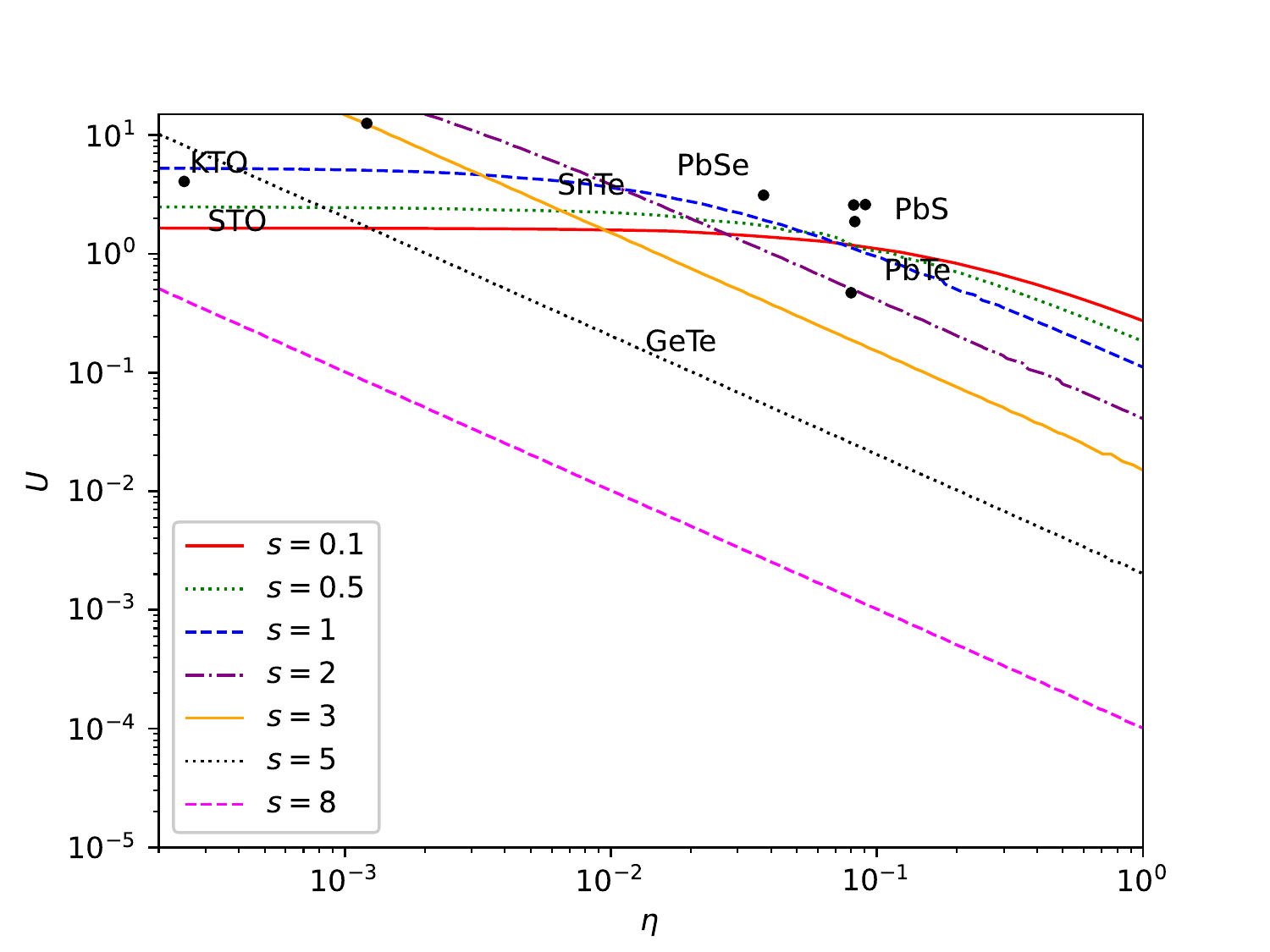}
    \caption{}
    \label{fig:nak_PD_zoom}
  \end{subfigure}
  \caption{a) Bipolaron binding energy $\Delta E = E_{opt} - E_{\infty}$ in units of the single polaron energy $|E_\infty| = 2|E_{pol}|$. Binding is found only in the strong-coupling regime where $U \gtrsim 16$. b) $\Delta E = 0$ curves at constant $\ts \equiv \sigma/l$ (see Section \ref{sec:methods}) in the weak-coupling regime show how binding can be achieved in situations where electron size is constrained to finite $\sigma$ - bipolaron formation is allowed below each $\ts$ curve. Known material parameters are plotted for comparison; STO refers to strontium titanate (SrTiO$_3$) and KTO refers to potassium tantalate (KTaO$_3$).}
  \label{fig:nak_PD}
\end{figure}

In this paper we develop a variational wave function that interpolates between the strong- and weak-coupling limits and captures bipolaron formation in a system of two isolated electrons, a suitable approximation for polar semiconductors in the anti-adiabatic, ultra-dilute limit. Our main result is shown in Fig.~\ref{fig:nak_PD} and described in Sections \ref{sec:results} and \ref{sec:discussion}. Although our wave function is able to capture two-electron correlations in the weak-coupling limit where it successfully reproduces known single-polaron physics, we find bipolaronic binding only in the strong-coupling regime. While the binding energy is $\sim10$\% of the single-polaron energy in some parts of the phase diagram, the transition to strong coupling becomes only slightly more favored than for the single polaron and occurs around $\alpha \sim 9$ for the smallest $\eta$, beyond the coupling strength of any materials we are aware of. It is possible, however, to induce bipolaron formation, including in weak coupling, by extrinsically constraining the size of the electron wave function. As shown in Fig. \ref{fig:nak_PD}b, the resulting phase diagram is controlled by the ratio of the wave function extent to an intrinsic phonon length scale. In all cases, our bipolaron wave function favors a finite electron separation, and curiously prefers to break rotational symmetry, although we cannot match the absolute magnitude of energy lowering obtained by wave functions with more variational parameters.

\section{Methods}\label{sec:methods}
We begin with the Fr{\"o}hlich Hamiltonian for two electrons of mass $m$ with positions $\Vec{r}_i$ and momenta $\Vec{p}_i$
\begin{align}
    \begin{split}\label{eqn:ham}
        H &= \sum_{i=1,2}\frac{\Vec{p}_i^2}{2m} + \sum_k \biggl \{ V_k b_k e^{i\Vec{k}\cdot \Vec{r}_i} + c.c. \biggr\} \\
        &+ \sum_k b_k^* b_k \hbar \omega_k + \frac{e^2}{\epsilon_\infty |\Vec{r}_1- \Vec{r}_2|},
    \end{split}
\end{align}
LO phonons with creation (annihilation) operators $b_k^*$ ($b_k$) at momentum $k$ are assumed to be dispersionless so that $\hbar\omega_k = \hbar \omega$. We also adopt the notation of LLP to define
\begin{equation}\label{eqn:V_k}
    V_k = -\frac{\hbar \omega i}{k}\left(\frac{\hbar}{2m\omega}\right)^{1/4}\left(\frac{4\pi\alpha}{V}\right)^{1/2} = -\frac{i2\hbar \omega }{k} \sqrt{\frac{\pi \alpha l}{V}},
\end{equation}
with a characteristic length scale $l \equiv \sqrt{\frac{\hbar}{2m \omega}}$ and Fr{\"o}hlich coupling constant
\begin{equation}
    \alpha = \frac{e^2}{\hbar}\left(\frac{1}{\epsilon_\infty}-\frac{1}{\epsilon_0}\right)\sqrt{\frac{m}{2\hbar \omega}} = (1-\eta)\frac{U}{2},
\end{equation}
where the Coulomb interaction strength $U = \frac{e^2/(\epsilon_\infty l)}{\hbar^2/(2ml^2)} = \frac{e^2}{\epsilon_\infty \hbar} \sqrt{\frac{2m}{\hbar \omega}}$ is the ratio of the effective Rydberg and phonon energies.

We propose a trial variational wave function
\begin{align}
    \begin{split}\label{eqn:wfn}
        \ket{\psi} &= e^{Q(\Vec{r}_1,\Vec{r}_2)} \psi_{el}(\Vec{r}_1,\Vec{r}_2) \ket{0}, \\
        e^{Q(\Vec{r}_1,\Vec{r}_2)} &= \exp\left[\sum_k (f_k b_k^* \Tilde{\rho}_k^* - f_k^* b_k \Tilde{\rho}_k) \right], \\
        \psi_{el}(\Vec{r}_1,\Vec{r}_2) &= \biggl( e^{-\frac{1}{2} \left( \frac{\Vec{r}_1-\Vec{\mu}_1}{\sigma}\right)^2} e^{-\frac{1}{2} \left( \frac{\Vec{r}_2-\Vec{\mu}_2}{\sigma}\right)^2} + (\Vec{r}_1 \to \Vec{r}_2) \biggr), \\
        \Tilde{\rho} &= e^{ia\Vec{k}\cdot \Vec{r}_1} + e^{ia\Vec{k}\cdot \Vec{r}_2},\quad b_k\ket{0} = 0
    \end{split}.
\end{align}
We treat the electrons as Gaussian wavepackets of characteristic size $\sigma$ centered at $\vec{\mu}_i$. As there are no spin operators in the Hamiltonian, we can choose to symmetrize or antisymmetrize the spatial wave function. Since the spin triplet configuration does not produce binding in the strong-coupling limit, we focus only on the spatially symmetric solution.

The coherent state $e^Q\ket{0}$ dresses the electrons with a semiclassical phonon distribution described by the variational function $f_k$. The form of $e^Q$ is adapted from the generalization of the LLP transformation introduced in Ref.~\cite{LDB77}. The addition of the variational parameter $a$, introduced in Ref.~\cite{Huy77}, allows us to tune between the weak- ($a=1$) and strong-coupling ($a=0$) limits. Physically, $a=1$ can be thought of as ``boosting'' the phonons into a frame instantaneously co-moving with the electron density. In Appendix \ref{app:nak_planewave} we verify that for uncorrelated plane wave electrons, applying this transformation with $a\to 1$ coincides with the results of LLP. The opposite $a=0$ limit centers the phonon distribution at a fixed point in space.

\begin{figure}[h!]
  \centering
  \begin{subfigure}{\linewidth}
    \includegraphics[width=0.5\linewidth]{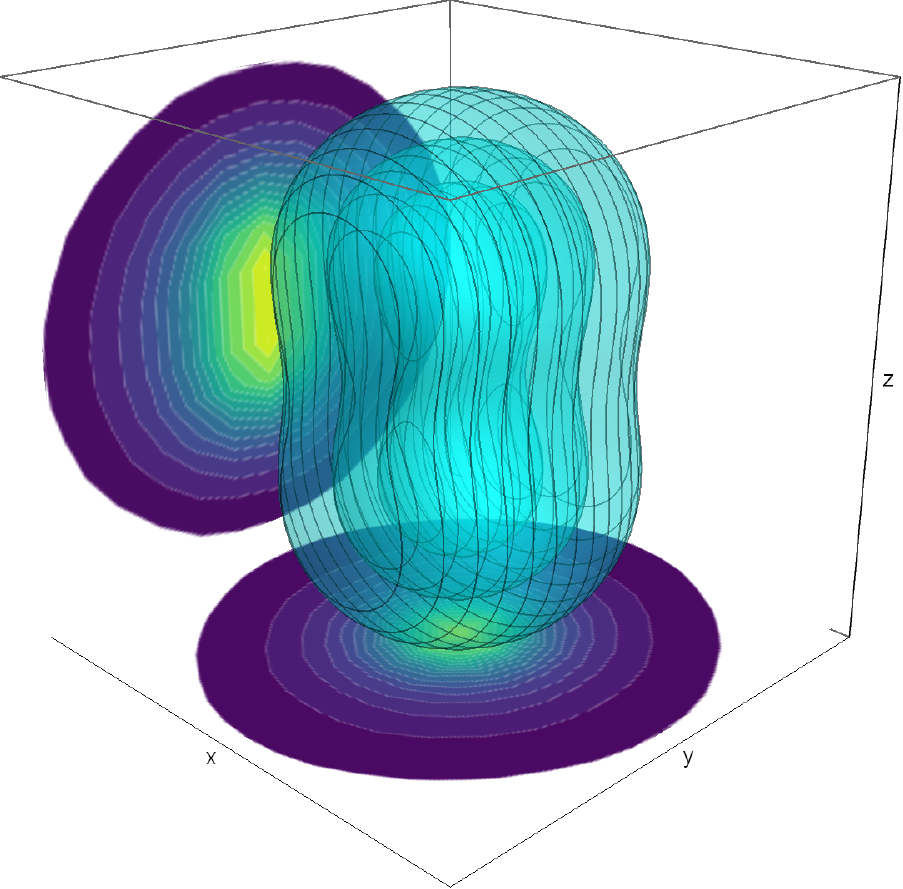}
    \caption{Strong coupling bipolaron solution}
    \label{fig:oblong}
  \end{subfigure}\\
  \begin{subfigure}{\linewidth}
    \includegraphics[width=0.5\linewidth]{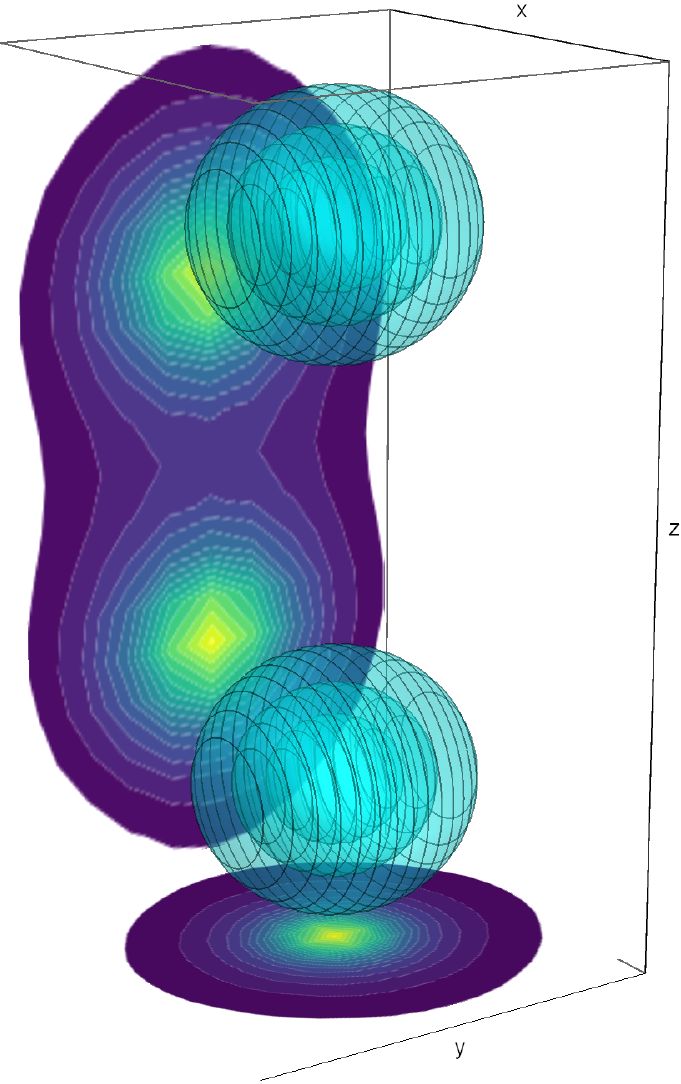}
    \caption{Well-separated electron limit}
    \label{fig:separated}
  \end{subfigure}
  \caption{3D contour plot in $x,y,z$ coordinates of the electron density at fixed $a=0.1$, for electron separation distance a) $d=2\sigma$ and b) $d=5\sigma$. The Fourier Transform of the phonon potential $f(\Vec{r})$ is projected onto the $xy$ and $yz$-planes of the boundary box.}
  \label{fig:nak_stages}
\end{figure}

We evaluate the energy expectation value $E = \bra{\psi}H\ket{\psi}$ and analytically solve $\delta E/\delta f^*_k = 0$ to obtain $f_k$, then numerically optimize the resulting expression as a function of the variational parameters $\ts \equiv \sigma/l$, $a$, and $y \equiv d/\sigma$. Further details of the calculation are provided in Appendix \ref{app:nakano}. Typical solutions are plotted in Fig. \ref{fig:nak_stages} as contours of constant electron density alongside the associated phonon displacements in the longitudinal direction.

We obtain the phase diagram in Fig. \ref{fig:nak_PD} by computing the energy difference between an optimized bipolaron solution and two well-separated polarons, $E(y\to \infty)$. In strong coupling with $a<1$, the numerical large-$y$ limit coincides with the single-polaron result in Appendix \ref{app:strong}. In weak coupling, the oscillations in the integrand at large $y$ produce large systematic errors on the order of one part in $10^3$, to which our phase diagram (Fig. \ref{fig:nak_PD_zoom}) is sensitive. We have verified that a numerical result with $y_\infty=10$, where the integration errors are acceptable, differs from $E_\infty = 2E_{\text{pol}}$ only by the semiclassical Coulomb energy difference, where $E_{\text{pol}}$ is the total energy of a single polaron.

\section{Results}\label{sec:results}

We first study the single-polaron properties of our wave function by taking the $y\to\infty$ limit where the Coulomb interaction between particles disappears. At small $\alpha$, the energy landscape as a function of $a$ has a single minimum at $a=1$. Here the wave function spreads out to minimize the kinetic energy and $\sigma \to \infty$, resulting in a linear polaronic energy lowering $-\alpha \hbar\omega$. As the coupling strength is increased, a metastable minimum around $a\sim 0.1$ develops and eventually becomes stable for $\alpha \sim 9$. After this point, the wave function tends to localize and produce an energy lowering quadratic in $\alpha$. In the very strong coupling limit this minimum evolves toward $a\to 0$. We expect that the discontinuous transition between strong and weak coupling is an artifact of our variational solution, despite allowing the possibility of $a$ varying continuously.

The inclusion of bipolaronic effects by allowing the energy to relax with respect to $y$ produces a regime of binding at strong coupling and small $\eta$, in qualitative agreement with Refs.~\cite{Devreese_2009, Lakhno10} and others, as shown in Fig. \ref{fig:nak_PD}, but does not substantially alter the rapid crossover from weak to strong coupling. The largest binding energies - up to around 13\% of the uncorrelated two-polaron energy - occur around the weakest allowed couplings, $(\eta, U) \sim (0, 20)$. As such the transition to strong coupling (from $a=1$ to $a<1$) occurs for the bipolaron at slightly smaller $\alpha$ than the analogous single polaron transition (compare $E_{opt}$ and $E_\infty$ in Supplementary Fig. \ref{fig:devcomp}).

\begin{figure}[h!]
  \centering
  \includegraphics[width=\linewidth]{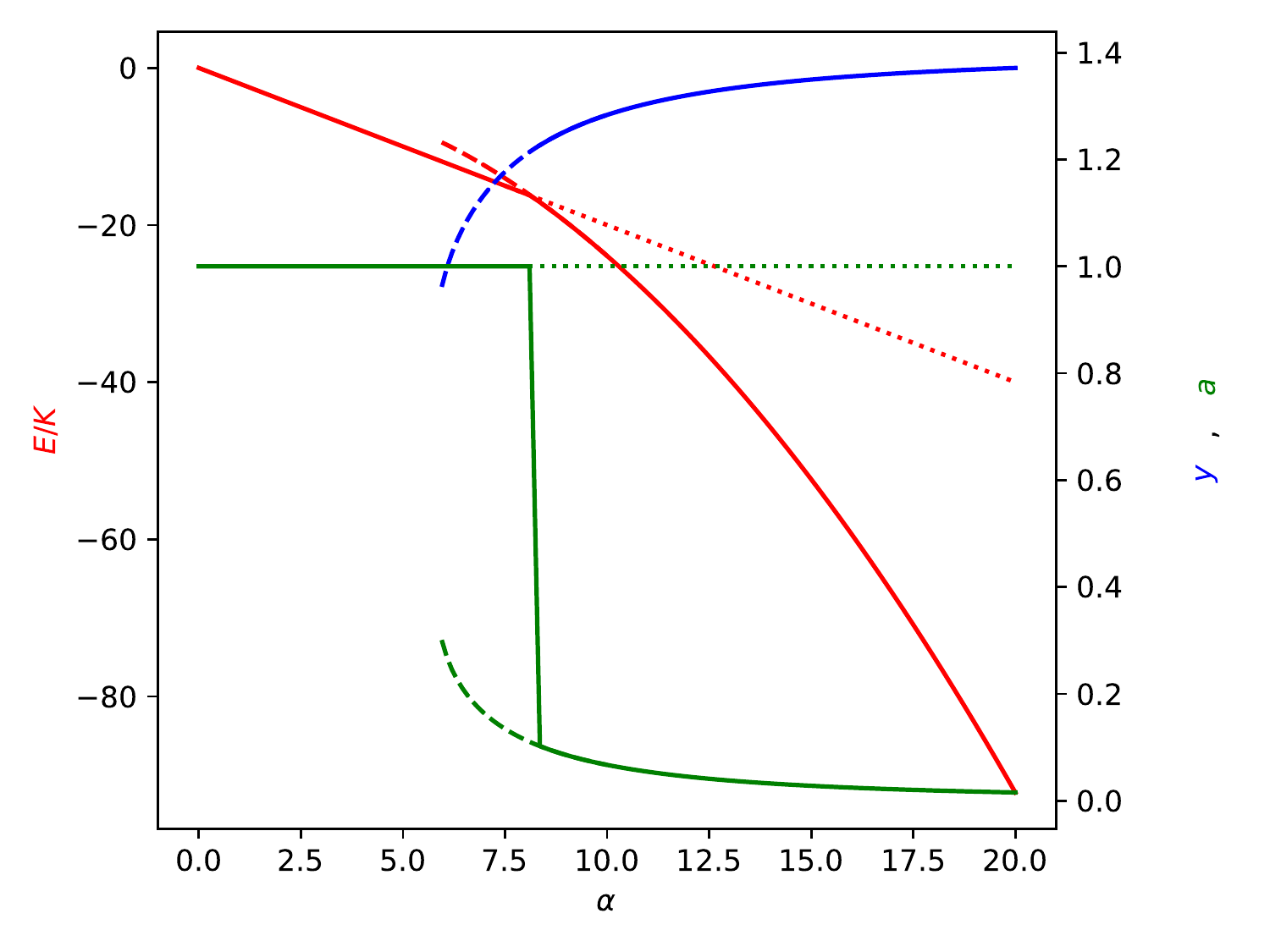}
    \caption{Energy, electron separation distance $y = d/\sigma$, and momentum coupling parameter $a$ plotted as a function of electron-phonon interaction strength $\alpha$, for ratio of dielectric constants $\eta = 0$ ($\eta > 0$ is qualitatively identical). Dashed lines indicate metastable strong-coupling solutions ($a < 0.5)$. Dotted lines denote a metastable $a=1$ minimum, corresponding to a plane wave solution. Only finite $y$ have been plotted.}
  \label{fig:firstorder}
\end{figure}

\begin{figure}[h!]
\begin{subfigure}{\linewidth}
  \centering
  \includegraphics[width=0.95\linewidth]{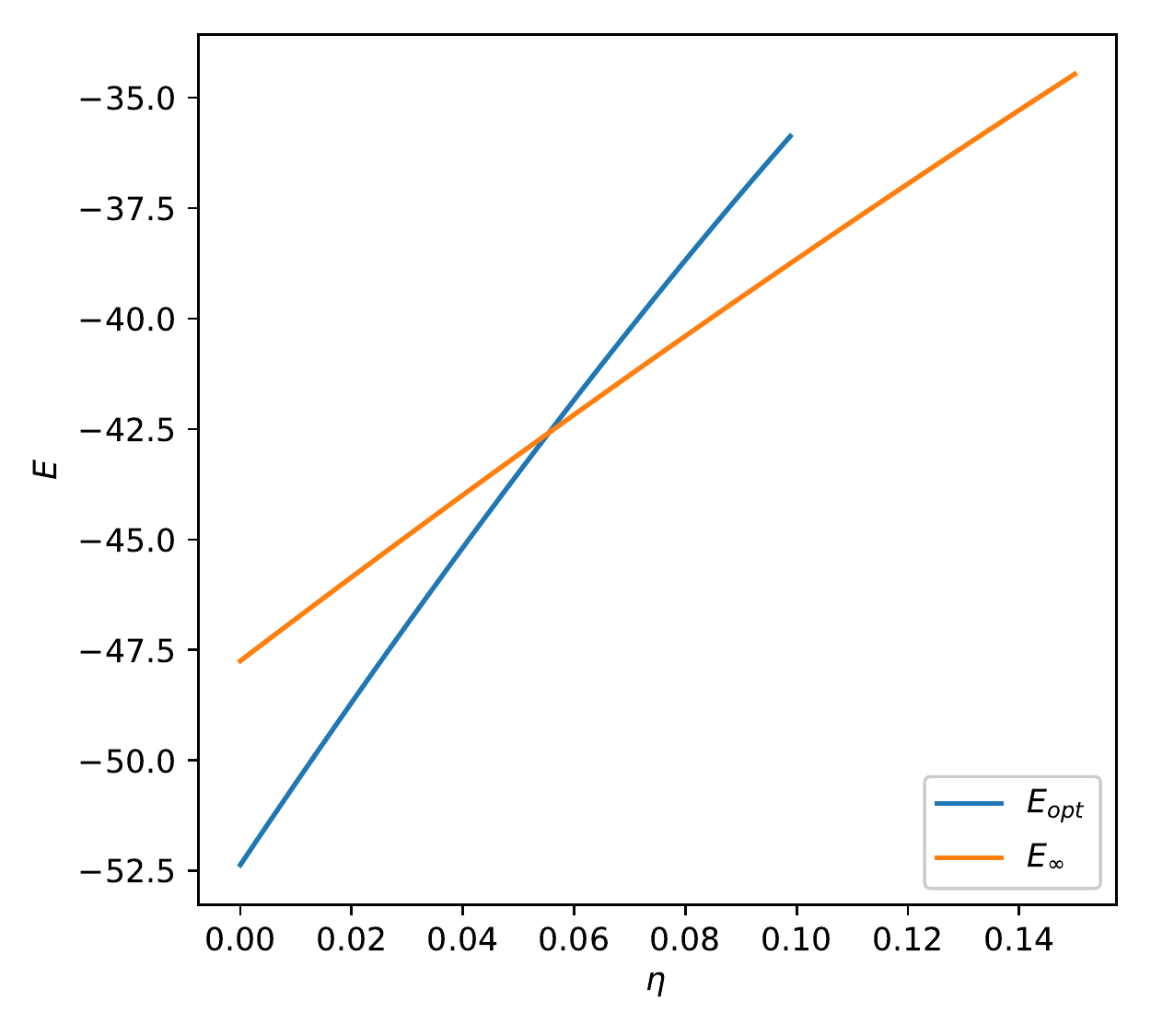}  
  \caption{}
  \label{fig:E(eta)}
\end{subfigure}\\
\begin{subfigure}{\linewidth}
  \centering
  \includegraphics[width=0.95\linewidth]{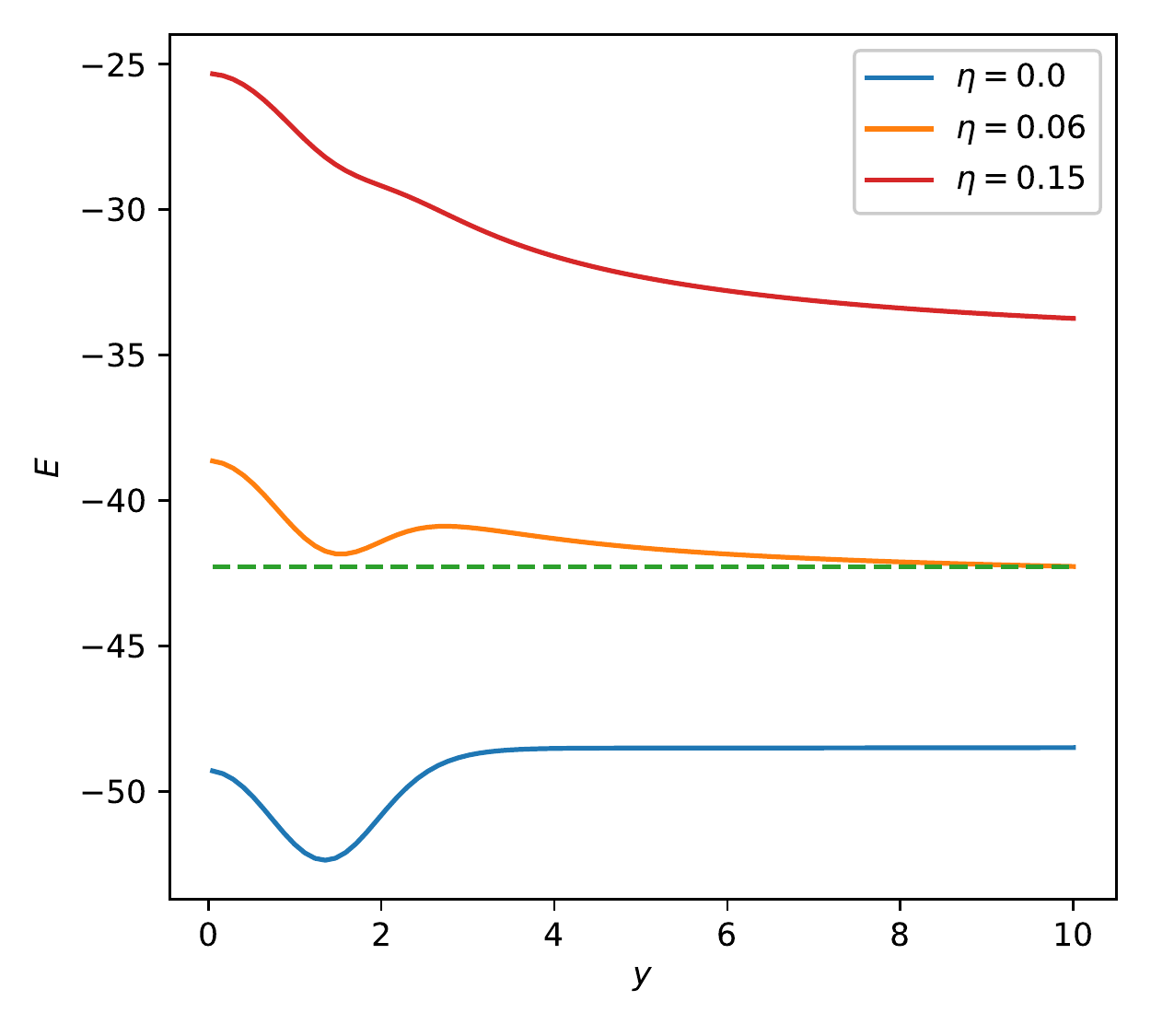}
  \caption{}
  \label{fig:E(y)}
\end{subfigure}
\caption{a) $E/K$ for fixed $U = 30$, with the bipolaron (blue) curve plotted for all $\eta$ where bipolaron formation exists. b) $E/K$ plotted as a function of electron separation distance for different values of $\eta$: for $\eta < \eta_c = 0.055$, we have stable bipolaron formation; for $\eta_c \lesssim \eta = 0.07$ we have a metastable state; and for $\eta = 0.2 \gg \eta_c$ there is no bipolaron formation.}
\label{fig:E_eta_y}
\end{figure}

We find the strongest binding for $y$ of order 1, with the optimal distance increasing at stronger coupling as shown in Fig. \ref{fig:firstorder}. A metastable bipolaron solution persists into smaller $U$ for some time, as shown by the dashed lines. Our bipolarons are also metastable as $\eta$ is increased, as shown in Fig. \ref{fig:E_eta_y} for $U=30$. In this regime the solution stays in strong coupling with finite $\sigma$, but Coulomb repulsion overwhelms the finite-$y$ bipolaron minimum.

The bipolaronic energy lowering within our formalism depends on the finite extent of the wave functions and is proportional to $1/\ts^2$ in the large-$\sigma$ limit, as explored in detail in Appendix \ref{app:nakano}. If $\sigma$ is extrinsically restricted (for instance due to device geometry or impurity effects), the wave function in weak coupling will still attempt to maximize its spatial extent, but a bipolaron minimum near $y=1$ may now exist where it is not overwhelmed by the Coulomb repulsion, which scales as $1/\ts$. In Fig. \ref{fig:nak_PD_zoom} we plot this weak-coupling phase boundary for different values of fixed $\ts$. Regions of allowed binding lie below each curve. Near $\eta = 1$, the phase boundaries fall off as $1/\eta$ and $1/\ts$, while at small $\eta$ they saturate at finite $U$. See Appendix \ref{app:beta} for details of the phenomenology and scaling at weak coupling.

\section{Discussion}\label{sec:discussion}
In the weak-coupling limit where $a=1$, the solution in the boosted frame can be understood semiclassically by studying the energy lowering of point charges in a dielectric medium, where the diverging result must be cut off by an electron radius $r_e$ (see Appendix \ref{app:classical} for details). The energies for one and two charges, respectively, are

\begin{align}
    \begin{split}\label{eqn:semiclassical}
        E_{\text{pol}} &= -\frac{e^2}{2\epsilon_\infty r_e} (1-\eta)= -\alpha \hbar \omega \\
        E_{\text{bi}}(d) &= -\frac{e^2}{\epsilon_\infty} (1-\eta) \left(\frac{1}{r_e} + \frac{1}{d} \right) + \frac{e^2}{\epsilon_\infty d} \\
        &= -2\alpha \hbar \omega + \frac{e^2}{\epsilon_0 d}
    \end{split}
\end{align}
where the single-polaron energy coincides with LLP when $r_e = l$. In the fully boosted $a=1$ frame, the phonons instantaneously follow the electron but quantum effects regularize a classically diverging phonon occupation. The two-polaron Coulomb interaction is screened down by $\eta$ but always remains repulsive, and therefore binding is never favored. Bipolaron formation is therefore a quantum effect made possible by the kinetic energy lowering of a molecular wave function compared to two separated polarons of finite extent, competing with the screened Coulomb repulsion.

Where bipolarons form, we have compared our wave function to a spherically symmetric Gaussian shell (see Appendix \ref{app:symm}) and find substantial energy lowering from breaking the symmetry, because in strong coupling the polaronic energy lowering is maximized by concentrating the charge density as tightly as possible. Our solutions therefore have a structure more similar to a hydrogen molecule than a helium atom. We obtain a further energy lowering by retaining a small but finite $a$, even at strong coupling. Despite these improvements, our solution is unable to match the energy lowering of those with more variational degrees of freedom, e.g. Ref.~\cite{devreese92}.
 
In the weak-coupling limit with constrained $\ts$, the bipolaron minimum always optimizes to $a=1$ and $y=1$ (within our ability to resolve), and the variational equations simplify considerably. The binding energy can be expressed as
\begin{equation}
    \Delta E = \frac{1}{\tilde\sigma^2}\Delta K +\frac{U}{\tilde\sigma}(1 -\beta(\ts))(1-\eta)\Delta V
    \label{eqn:dE}
\end{equation}
where $\Delta K < 0$ is the kinetic energy difference between the bipolaron wave function and two well-separated polarons, and $\Delta V$ is the Coulomb integral for $y=1$. In the semiclassical large-$\ts$ limit, $\beta(\ts) = 0$ and the second term in (\ref{eqn:dE}) reduces to the screened Coulomb repulsion, while elsewhere $\beta$ encodes the quantum effects that cause the phonons to under-screen on length scales $<l$. $\Delta E = 0$ can be solved analytically and for small $\beta$ and $\eta$ gives a phase boundary $U\sim 1/(\eta\ts)$, in agreement with Fig. \ref{fig:nak_PD_zoom}. In Appendix \ref{app:beta} we show that at large $\ts$, the leading-order contribution to $\beta$ is $1/\ts^2$, while numerically we find $\beta \sim .1$ at small $\ts$, causing the phase boundary to bend away from $1/\eta$ behavior and saturate at finite $U$. This produces two distinct regimes of weak-coupling bipolaron formation: at small $\ts$ the kinetic energy term in (\ref{eqn:dE}) is maximized, but the growth in $\beta$ eventually outcompetes it and the phase boundary plateaus completely near $U=1$. On the other hand, it is always possible at larger $\ts$ to avoid this effect by staying deeper in the semiclassical limit of the phonons; although this pushes the phase boundary to smaller $U$, it is now possible to access new regimes of the phase diagram at large $U$ by reducing the screened Coulomb repulsion (decreasing $\eta$). 

\subsubsection{Bipolarons in real materials}
We plot several $(\eta, U)$ values of materials which could host bipolarons on our phase diagram in Fig. \ref{fig:nak_PD}. All the candidates shown lie within the weak-coupling region, in which we observe no binding. However, the near-ferroelectric perovskites \ce{SrTiO3} (STO) and \ce{KTaO3} (KTO) lie in the small-$\eta$ regime where binding is possible if $\sigma$ is extrinsically restricted. Both of these materials are unconventional dilute superconductors when doped, and bipolarons are a candidate for a mechanism of preformed pair formation \cite{Collignon19,Swartz18,Terry19,Lin13,Liu716}. In Ref.~\cite{Bussmann-Holder}, metallic filaments of approximately 60nm diameter were measured in oxygen-reduced STO, while in Ref.~\cite{Cheng15} evidence of preformed pairing beyond 900mK was observed in nanowires 5nm wide fabricated by oxygen reduction in a lanthanum aluminate-STO heterostructure. Using the parameters of bulk STO, $(\eta_{STO}, U_{STO}) = (2\times 10^{-4}, 4)$, these confined geometries correspond to $\ts$ of 97 and 8.1, respectively, from which we obtain binding energies of $1.3\ \mathrm{\mu eV}$ and $.41\ \mathrm{meV}$. Using $\Delta E \sim k_BT$, these would correspond to pairing temperatures of 0.016 K and 4.8 K. The latter is well above STO's optimal superconducting transition temperature $T_c \sim 0.3\ \mathrm{K}$. Such a mechanism of preformed pair formation could explain the observation of these phenomena only in confined geometries, although this scenario would have to be justified by microscopic calculations.

\section{Conclusions}\label{sec:conclusions}
In this paper we have introduced a variational method to calculate bipolaron binding in the Fr{\"o}hlich model, that is valid in both strong- and weak-coupling limits. Our strong-coupling results are consistent with previous work, but we find the crossover to weak coupling essentially destroys the bipolaronic binding at moderate values of coupling, except as a metastable state. However, we find that confinement may stabilise bipolarons in a regime of parameters which may be relevant for nearly ferro-electric oxides. We find that bipolarons, where they exist, are generally anisotropic - i.e. a molecular ``\ce{H2}-like'' configuration, rather than a spherically symmetric ``\ce{He}-like'' arrangement.

Though our assumption that the dominant coupling is to the LO phonon mode is empirically justified for materials such as STO \cite{Devreese10}, our use of the Fr{\"o}hlich Hamiltonian neglects short-range interactions, in particular deformation potentials associated with acoustic phonons. Relatedly, the lack of binding we find below around $\alpha \sim 9$ for unconstrained $\sigma_{max}$ could be due to our inclusion of only a single phonon mode, whereas for instance STO has not one but three LO modes; \cite{Devreese92b} indicates that interactions between electrons and multiple phonon branches could increase the region of bipolaron stability. 

Our calculations are limited by the use of a semiclassical coherent state phonon wave function. Quantum Monte Carlo (QMC) simulations would allow us to incorporate higher-order phonon correlations, which have been shown to be important in the intermediate-coupling regime for the single polaron problem \cite{Svistunov2000}. 

Our calculations also assume a two-particle system, and thus are applicable only to the extremely dilute regime where the bipolaron size is much smaller than the inter-electron spacing. If the rotational symmetry breaking we observe here persists in the many-body context, it could favor electronic nematic states. The interplay of bipolaron formation and Wigner crystallization, with possible intermediate tripolaron or other multi-polaron regimes, could also be studied by QMC simulations.

\begin{acknowledgments}
L. L. was supported by the University of Chicago CCRF College Research Fellows Program and Dean's Fund Award. Work at Argonne was supported by DOE, Office of Science, Basic Energy Sciences, Materials Science and Engineering under Contract No. DE-AC02-06CH11357.
\end{acknowledgments}

\appendix
\section{Semiclassical analysis} \label{app:classical}
The Lee-Low-Pines (LLP) analysis of a single polaron can be presented semiclassically - a classical analysis of a point charge in a dielectric medium with the short-range (UV) singularity regularised by quantum fluctuations. This viewpoint provides considerable intuition as to how to deal with the bipolaron problem.

Imagine that we drop a lone electron into an infinite dielectric, i.e. an elastic, polarizable medium. We want to find the energy cost associated with letting the medium polarize. Since the electron is the only charge in the dielectric, any polarization within the material will be caused solely by the electron's field. Dipoles in the dielectric can be modeled by (linearly) polarizable degrees of freedom. 

Let's first find the E-field generated by this single free (point) charge. From Maxwell's equations (in Gaussian units),
\begin{equation}
    \Vec{D} = \Vec{E} + 4\pi \Vec{P}.
\end{equation}
We want to write a functional for the total energy of the system. However, we want to avoid self-interaction terms of the electron with its own electric field. The electron doesn't feel the field it itself generates, so there's no work associated with moving it through its own field. As such, the relevant E-field is only the one induced by the polarization of the dipoles in the material,
\begin{equation}
    \Vec{E} = -4\pi \Vec{P}
\end{equation}

The dipole moment of a single oscillator is
\begin{equation}
    \Vec{p} = Z^* \Vec{u}(\Vec{r}),
\end{equation}
where $Z^*$ is the effective charge displaced in a single unit cell by its displacement $\Vec{u}$ from equilibrium. By definition, the polarization of the whole system is the dipole moment per unit volume, i.e. $\Vec{p} = \Vec{P} \Delta V$, where $\Delta V$ is the volume of one unit cell. Thus
\begin{equation}
    \Vec{P} = \frac{Z^*}{\Delta V} \Vec{u}(\Vec{r}) = \rho^* \Vec{u}(\Vec{r}),
\end{equation}
where $\rho^* \equiv Z^*/\Delta V$ is the effective charge per unit cell, and so the electric field is
\begin{equation}
    \Vec{E} = -4\pi \rho^* \Vec{u}(\Vec{r}).
\end{equation}
The electric potential is
\begin{equation}
    V(\Vec{r}) = \int \frac{\Vec{P}(\Vec{r}') \cdot \hat{\tr}}{\tr^2} d\tau',
\end{equation}
where $\Vec{\tr}_i = \Vec{r}_i - \Vec{r}'$ and integration is performed over the primed coordinates. Then the energy functional is
\begin{align}
    \begin{split}\label{eqn:E_real_fxnal_1e}
        E[\Vec{u}(\Vec{r})] &= -eV(\Vec{r}) + \frac{1}{2} \rho_M \omega^2 \int |\Vec{u}(\Vec{r}')|^2 d^3\Vec{r}' \\
        &= -e\rho^* \int \frac{ \Vec{u}(\Vec{r}') \cdot \hat{\tr}}{\tr^2} d^3\Vec{r}' + \frac{1}{2} \rho_M \omega^2 \int |\Vec{u}(\Vec{r}')|^2 d^3\Vec{r}' \\
    \end{split}
\end{align}
Minimizing our Lagrangian 
\begin{equation}
    L =  -e\rho^* \frac{ \Vec{u}(\Vec{r}') \cdot \hat{\tr}}{\tr^2} + \frac{1}{2} \rho_M \omega^2 |\Vec{u}(\Vec{r}')|^2
\end{equation}
with respect to $\Vec{u}(\Vec{r}')$,
\begin{equation}
    \Vec{u}(\Vec{r}') = \frac{e\rho^*}{\rho_M \omega^2} \frac{\hat{\tr}}{|\Vec{r}-\Vec{r}'|^2}
\end{equation}
We now plug this result back in and solve for the energy:
\begin{align}
    \begin{split}
        E &= -\frac{e^2\rho^{*2}}{2\rho_M \omega^2} \int \frac{1}{|\Vec{r}-\Vec{r}'|^4} d^3\Vec{r}' \\
        &= -\frac{2\pi e^2\rho^{*2}}{ \rho_M \omega^2 r_e}
    \end{split}
\end{align}
which diverges as $r_e \rightarrow 0$ for a classical point electron. However, in actuality $r_e$ will be cut off by some small, finite length scale, e.g. the size of the electron, lattice spacing, etc., and in a continuum model the natural cutoff is the phonon length $l$ on which the electron density is smeared.

\subsubsection{Comparison with LLP energy shift}
We use the following two relations to convert the above expression into LLP units:
\begin{align}
    \begin{split}\label{eqn:omegas}
        \omega_L^2-\omega_T^2 &= \frac{4\pi \epsilon_\infty \rho^{*2}}{\rho_M} \\
        \frac{\omega_L^2}{\omega_T^2} &= \frac{\epsilon_{sr}}{\epsilon_\infty}, \\
    \end{split}
\end{align}
where $\omega_L = \omega$ and $\omega_T$ are the longitudinal and transverse optic phonon frequencies of the phonons in the dielectric, and $\epsilon_{sr}$ refers to the (dimensionless) static relative dielectric constant (i.e. at zero frequency). Combining the two equations above, we obtain
\begin{equation}
    \frac{1}{\epsilon_\infty} - \frac{1}{\epsilon_{sr}} = \frac{4\pi \rho^{*2}}{\rho_M \omega_L^2}
\end{equation}

Our quantity of comparison is the constant energy shift due to one polaron, $\Delta E = -\alpha \hbar \omega$. Using Eqns \ref{eqn:omegas} and plugging in the phonon length scale for our integration cutoff, i.e. $r_e = l = \sqrt{\frac{\hbar}{2m\omega}}$, we find that 
\begin{equation}
    E_{Clas} = -\frac{e^2}{\sqrt{2}}  \biggl( \frac{1}{\epsilon_\infty} -\frac{1}{\epsilon_{sr}} \biggr) \sqrt{\frac{m\omega}{\hbar}}
\end{equation}
and
\begin{equation}
    E_{LLP} = -\frac{e^2}{\sqrt{2}}\biggl( \frac{1}{\epsilon_\infty}-\frac{1}{\epsilon_{sr}} \biggr) \sqrt{\frac{m\omega}{\hbar}}
\end{equation}
The LLP methodology can thus be explained straightforwardly by semiclassical physics. Extending the above strategy to the two electron case gives us equation \ref{eqn:semiclassical}.

\section{Strong coupling}\label{app:strong}
\begin{figure}
    \centering
    \includegraphics[width=\linewidth]{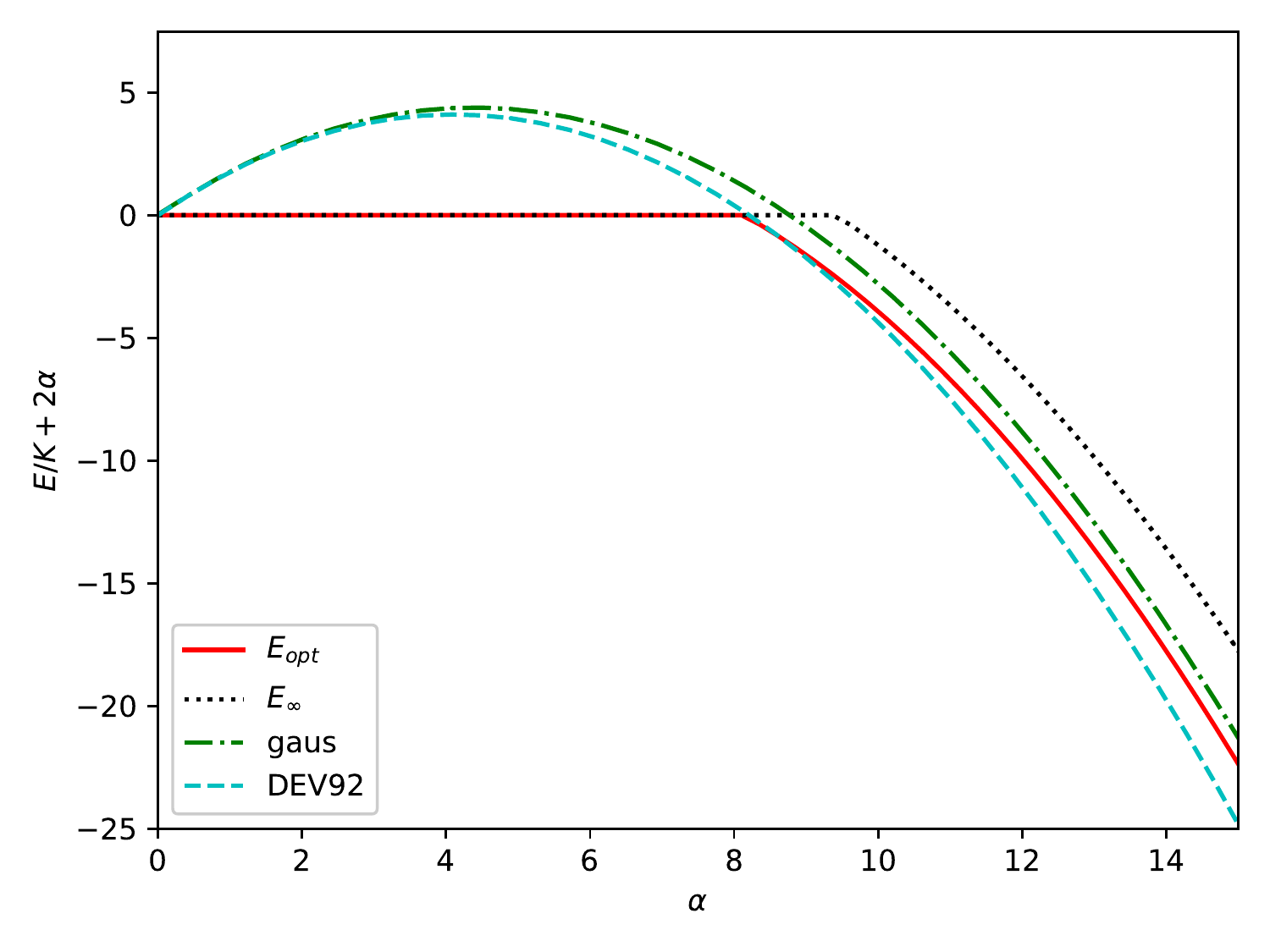}
    \caption{Optimized bipolaron and separated polaron energies $E_{opt}$ and $E_\infty$ plotted as a function of electron-phonon coupling strength $\alpha$. `DEV92' refers to the oscillator-oscillator wave function presented in \cite{devreese92}; it is identical to the wave function in equation \ref{eqn:gaussE} except that it allows an extra degree of freedom in the electron size $\sigma$. `$a=0$' refers to the Gaussian strong coupling wave function described in Appendix \ref{app:strong}.}
   \label{fig:devcomp}
\end{figure}

In the strong-coupling limit, we use the variational trial wave function \begin{equation} \label{eqn:gaus}
    \ket{\psi} = \psi_{el} U_2 \ket{0},
\end{equation}
where
\begin{align}
    \begin{split}
        \psi_{el}(\Vec{r}_1, \Vec{r}_2) &= A \biggl( e^{-\frac{1}{2} \left( \frac{\Vec{r}_1-\Vec{\mu}_1}{\sigma}\right)^2} e^{-\frac{1}{2} \left( \frac{\Vec{r}_2-\Vec{\mu}_2}{\sigma} \right)^2} \\
        &+ e^{-\frac{1}{2} \left( \frac{\Vec{r}_1-\Vec{\mu}_2}{\sigma}\right)^2} e^{-\frac{1}{2} \left( \frac{\Vec{r}_2-\Vec{\mu}_1}{\sigma}\right)^2} \biggr) \\
        A^2 &= \frac{1}{2 \pi^3 \sigma^6 (1 + e^{\frac{-d^2}{2\sigma^2}})} \\
        U_2 &= \prod_k e^{b_k^* f(\Vec{k}) - b_k f^*(\Vec{k})}
    \end{split}
\end{align}
Note that this is identical to our original wave function (equation \ref{eqn:wfn}), taking $a \to 0$. Following the procedure in Section \ref{sec:methods}, we obtain as our ground state energy estimate
\begin{align}
    \begin{split}\label{eqn:gaussE}
        E_{\text{gaus}} &= \frac{1}{\ts^2} \biggl[ 3 - \Delta K\biggr] - \frac{U(1-\eta)}{\ts (1  + e^{-\frac{y^2}{2}})^2} \sqrt{\frac{2}{\pi}} \biggl[ 1 +2 e^{-y^2} \\
        &+ \frac{1}{y} \sqrt{\frac{\pi}{2}} \biggl( \Erf\left( \frac{y}{ \sqrt{2}} \right) + 8 e^{-\frac{y^2}{2}} \Erf\left( \frac{y}{ 2\sqrt{2}} \right) \biggr) \biggr] \\
        &+ \frac{U}{\ts} 
        \Delta V\\
        \Delta K &= \frac{y^2/2}{e^{\frac{y^2}{2}} + 1} \\
        \Delta V &= \frac{ \frac{1}{y} \Erf \bigl( \frac{y}{\sqrt{2} } \bigr) + \sqrt{\frac{2}{\pi}} e^{-\frac{y^2}{2}}}{1 + e^{-\frac{y^2}{2}}}
    \end{split}
\end{align}
where $y=d/\sigma$ and $\ts = \sigma/l$, and the energy is in units of the kinetic energy $\frac{\hbar^2}{2ml^2} = \hbar \omega$. In the large $y$ limit, 
\begin{align}
    \begin{split} \label{eqn:gaus_lgy}
        \sigma_{\text{min}} &= \frac{3\hbar}{2\alpha m\omega l} \sqrt{\frac{\pi}{2}} = \frac{3l}{\alpha} \sqrt{\frac{\pi}{2}} = \frac{3\hbar^2 \epsilon_\infty}{m e^2 (1-\eta)} \sqrt{\frac{\pi}{2}} \\
        E_{\text{min}} &= -\frac{2 \alpha^2 \hbar \omega}{3\pi}
    \end{split}
\end{align}
For the single polaron, we find an energy lowering of $-\alpha^2 \hbar \omega/(3\pi)$, identical to the (leading order) strong-coupling results of \cite{devreese92}, \cite{feyn55}, and \cite{Huy77}. This solution is plotted alongside the others discussed in Fig. \ref{fig:devcomp}. 

\subsubsection{Radially symmetric wave function}\label{app:symm}
In the strong coupling limit, we also find it of interest to compare our result in the next section with a radially symmetric wave function to determine whether a symmetry-breaking wave function is preferred by this model/system. We thus begin with the spherically symmetric wave function
\begin{align}
    \begin{split}\label{eqn:nak_psi_symm}
        \psi(\Vec{r}) &= A e^{-\frac{(|\Vec{r}|- |\Vec{d}|)^2}{4\sigma^2}} \\
        \psi(\Vec{R}) &= \left( \frac{2}{\pi \sigma^2} \right)^{3/4} e^{-(\frac{\Vec{R} -\Vec{M}}{\sigma})^2}
    \end{split}
\end{align}
with normalization constant
\begin{equation}
    |A| = (2\pi \sigma^2)^{-3/4} \biggl[ \sqrt{\frac 2\pi} ye^{-y^2} + (1+y^2) \biggl(1 + \Erf \biggl(\frac{y}{\sqrt{2}} \biggr) \biggr) \biggr]^{-1/2}
\end{equation}
The Hamiltonian in equation \ref{eqn:ham}, written in relative ($\vec r = \vec r_1 - \vec r_2,\, \vec p = \vec p_1 - \vec p_2$) and center-of-mass ($\vec R = (\vec r_1 + \vec r_2)/2, \vec P = \vec p_1 + \vec p_2$) coordinates is
\begin{align}
    \begin{split}
        H &= \frac{p^2}{4m}+\frac{P^2}{4m}+\sum_k \biggl\{ V_k a_k e^{i\Vec{k} \cdot \Vec{R}} (e^{i\Vec{k} \cdot \Vec{r}/2} + e^{-i\Vec{k} \cdot \Vec{r}/2}) + c.c. \biggr \} \\
        &+ \sum_k a_k^* a_k \hbar \omega + \frac{e^2}{\epsilon_\infty r}.
    \end{split}
\end{align}

\begin{figure}
    \centering
    \includegraphics[width=0.9\linewidth]{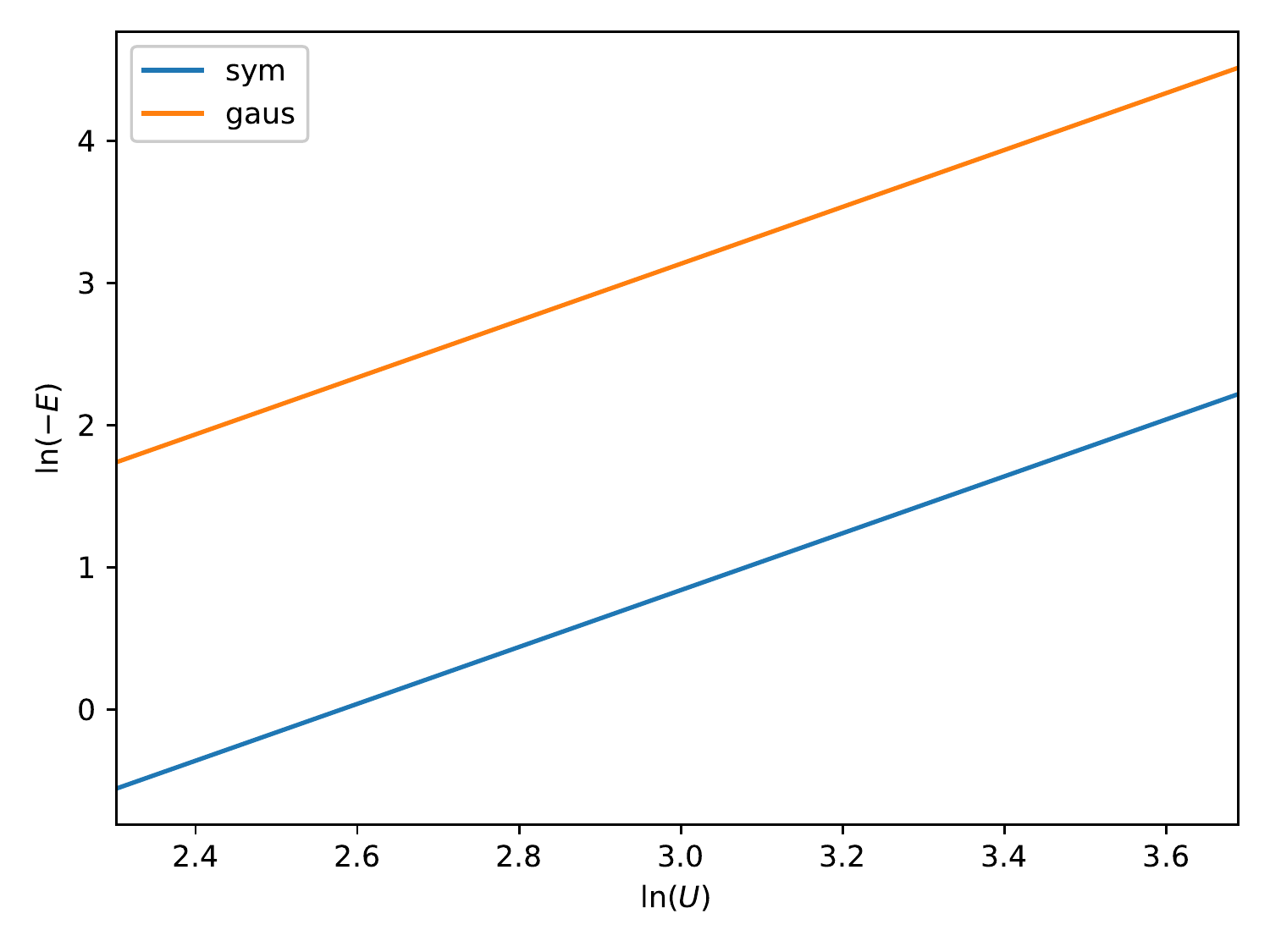}
    \caption{Comparison between bipolaron energies ($E_{opt}$) obtained from spherically symmetric (see equation \ref{eqn:nak_psi_symm}; orange) and our strong coupling (symmetry breaking, see equation \ref{eqn:gaus}; blue) wave functions.}
    \label{fig:sym_energy_comp}
\end{figure}

Its expectation value is then
\begin{align}
    \begin{split}
        \braket{H} &= \frac{\hbar^2}{4m\sigma^2} \biggl[ \frac{3 + 5y^2 + \sqrt{\frac 2\pi} y e^{-y^2} + (y-1)(y-3) \Erf (\frac{y}{\sqrt 2})}{\sqrt{\frac 2\pi} ye^{-y^2} + (1+y^2) \left(1 + \Erf (\frac{y}{\sqrt{2}}) \right)} \\
        &+ 3 \biggr] + \frac{e^2}{\epsilon_\infty \sigma} \frac{e^{-y^2} + (1+y) \Erf (\frac{y}{\sqrt 2})}{\sqrt{\frac 2\pi} ye^{-y^2} + (1+y^2) \left(1 + \Erf (\frac{y}{\sqrt{2}} ) \right)} \\
        &\sum_k |f_k|^2 \hbar \omega + 2 \sum_k (V_k f_k + V_k^* f_k^*) e^{-k^2\sigma^2/8} \frac{8\pi |A|^2}{k} I
    \end{split}
\end{align}
where we have denoted the integral
\begin{equation}
    I \equiv \int_0^\infty dr\, r e^{-(r - d)^2/\sigma^2} \sin (kr/2)
\end{equation}
\comment{Solving for $f_k$, we obtain
\begin{equation}
    f_k = \frac{-2V_k^* e^{-k^2\sigma^2/8} \frac{8\pi}{k}|A|^2 I}{\hbar \omega}
\end{equation}
Plugging back in, our Hamiltonian becomes
\begin{align}
    \begin{split}
        \braket{H} &= \frac{\hbar^2}{4m\sigma^2} \biggl[ \frac{3 + 5y^2 + \sqrt{\frac 2\pi} y e^{-y^2} + (y-1)(y-3) \Erf (\frac{y}{\sqrt 2})}{\sqrt{\frac 2\pi} ye^{-y^2} + (1+y^2) \left(1 + \Erf (\frac{y}{\sqrt{2}}) \right)} \\
        &+ 3 \biggr] - \frac{4 (8\pi |A|^2)^2}{\hbar \omega} \sum_k \frac{|V_k|^2 e^{-k^2\sigma^2/4} I^2}{k^2} \\
        &+ \frac{e^2}{\epsilon_\infty \sigma} \frac{e^{-y^2} + (1+y) \Erf (\frac{y}{\sqrt 2})}{\sqrt{\frac 2\pi} ye^{-y^2} + (1+y^2) \left(1 + \Erf (\frac{y}{\sqrt{2}} ) \right)}
    \end{split}
\end{align}
Converting the k-sum into an integral, we have 
\begin{align}
    \begin{split}
        \braket{H} &= \frac{\hbar^2}{4m\sigma^2} \biggl[ \frac{3 + 5y^2 + \sqrt{\frac 2\pi} y e^{-y^2} + (y-1)(y-3) \Erf (\frac{y}{\sqrt 2})}{\sqrt{\frac 2\pi} ye^{-y^2} + (1+y^2) \left(1 + \Erf (\frac{y}{\sqrt{2}}) \right)} \\
        &+ 3 \biggr] - 512\pi \alpha \hbar \omega l |A|^4 \int_0^\infty dk \frac{e^{-k^2\sigma^2/4} I^2}{k^2} \\
        &+ \frac{e^2}{\epsilon_\infty \sigma} \frac{e^{-y^2} + (1+y) \Erf (\frac{y}{\sqrt 2})}{\sqrt{\frac 2\pi} ye^{-y^2} + (1+y^2) \left(1 + \Erf (\frac{y}{\sqrt{2}} ) \right)}
    \end{split}
\end{align}
Changing integration variables to $r' = r/\sigma$, $k' = k\sigma$ and writing the energy in units of $\hbar \omega$,
}
Following the procedure in Section \ref{sec:methods} and nondimensionalizing, we obtain an energy
\begin{align}
    \begin{split}
        \braket{H} &= \frac{1}{2\ts^2} \biggl[ \frac{3 + 5y^2 + \sqrt{\frac 2\pi} y e^{-y^2} + (y-1)(y-3) \Erf (\frac{y}{\sqrt 2})}{\sqrt{\frac 2\pi} ye^{-y^2} + (1+y^2) \left(1 + \Erf (\frac{y}{\sqrt{2}}) \right)} \\
        &+ 3 \biggr] + \frac{U}{\ts} \frac{e^{-y^2} + (1+y) \Erf (\frac{y}{\sqrt 2})}{\sqrt{\frac 2\pi} ye^{-y^2} + (1+y^2) \left(1 + \Erf (\frac{y}{\sqrt{2}} ) \right)}\\
        &-\frac{64 \alpha}{\pi^2 (\sqrt{\frac 2\pi} ye^{-y^2} + (1+y^2) (1 + \Erf (\frac{y}{\sqrt{2}})))^2} \frac{1}{\ts} \\
        &*\int_0^\infty \frac{dk' e^{-(k')^2/4}}{(k')^2} \biggl( \int_0^\infty dr'\, r' e^{-(r' - y)^2} \sin (k'r'/2) \biggr)^2 
    \end{split}
\end{align}
which must be evaluated numerically, but is structurally similar to the other strong-coupling solutions we have encountered. A comparison between the optimized bipolaron energies of this spherically symmetric wave function and our symmetry-breaking solution in strong coupling is shown in Fig. \ref{fig:sym_energy_comp}.

\section{Generalizing the Lee-Low-Pines Transform}\label{app:nak_planewave}
Lee, Low, and Pines' formulation of the single polaron problem is in essence a variational wave function which is an eigenstate of the total momentum $\vec{P}$:
\begin{align}
    \begin{split}
        \ket{\vec{r},\vec{P}} &= e^{\frac{i}{\hbar}\vec{P} \cdot \vec{r}} ST \ket{0} \\
        S &= e^{-i\sum_k b_k^* b_k \vec{k} \cdot \vec{r}} \\
        T &= e^{\sum_k(b_k^* f_k - b_k f^*_k)}
    \end{split}
\end{align}
where $S$ and $U$ are unitary transformations; $S$ couples the phonon momenta with the electron position and $U$ is a coherent state with momentum amplitudes $f_k$. We can use the fact that $S\ket{0} = \ket{0}$ to rewrite
\begin{align}
    \begin{split}
        \ket{\vec{r},\vec{P}} &= e^{\frac{i}{\hbar}\vec{P} \cdot \vec{r}} ST \ket{0} = e^{\frac{i}{\hbar}\vec{P} \cdot \vec{r}} ST S^{-1} S \ket{0} \\
        &= e^{\frac{i}{\hbar}\vec{P} \cdot \vec{r}} US \ket{0} = e^{\frac{i}{\hbar}\vec{P} \cdot \vec{r}} U \ket{0}
    \end{split}
\end{align}
where 
\begin{equation}
    U \equiv STS^{-1} \equiv e^{Q(\vec{r})} = \exp\left[\sum_k (e^{i\vec{k} \cdot \vec{r}} f_k b_k^* - e^{i\vec{k} \cdot \vec{r}} f_k^* b_k \right]
\end{equation}
The many-body generalization of the above $U$ transformation is given in \cite{Nakano16} and \cite{LDB77} as
\begin{equation}
    Q(\vec{r}) \to \sum_i Q(\vec{r}_i) = \int_r \hat{n}(\vec{r}) Q(\vec{r}),
\end{equation}
where $\hat{n}$ is the fermion density operator and $\int_r$ denotes integration over all space. For the two-electron problem at hand, 
\begin{equation}
    e^{Q(\Vec{r}_1,\Vec{r}_2)} = \exp\left[\sum_k (f_k b_k^* \Tilde{\rho}_k^* - f_k^* b_k \Tilde{\rho}_k) \right],
\end{equation}
where $\Tilde{\rho} = e^{ia\Vec{k}\cdot \Vec{r}_1} + e^{ia\Vec{k}\cdot \Vec{r}_2}$ is the modified electron density incorporating the phonon momentum ``boost'' parameter used by \cite{Huy77} to interpolate between the weak- ($a\to 1$) and strong-coupling ($a\to 0$) limits.

To check the validity of this somewhat unintuitive transformation, we apply it to a two-electron plane wave product state confined to a box of size $V = L^3$:
\begin{equation}
    \ket{\psi} = U \frac{e^{i\vec{q} \cdot (\Vec{r}_1 + \vec{r}_2)}}{V} \ket{0}
\end{equation}
which gives us the following operator transformation rules:
\begin{align}
    \begin{split}
        U^{-1} \hat{p}_i U &= \hat{p}_i - a\sum \hbar \Vec{k} (e^{-iak\cdot r_i} f_k b_k^* + f_k^* e^{iak\cdot r_i} b_k) \\
        &-a \sum_k \hbar \Vec{k} |f_k|^2 (1+ \cos(a\Vec{k} \cdot \Vec{r})) \\
        U^{-1} b_k U &= b_k + f_k (e^{-ia\Vec{k} \cdot \Vec{r}_1} + e^{-ia\Vec{k} \cdot \Vec{r}_2}) \\
        U^{-1} b_k^* U &= b_k^* + f_k^* (e^{ia\Vec{k} \cdot \Vec{r}_1} + e^{ia\Vec{k} \cdot \Vec{r}_2})
    \end{split}
\end{align}
where $\vec{r} = \vec{r}_1 - \vec{r}_2$. Applying our periodic boundary condition $e^{iqL} = e^{ikL} = 1$, we obtain the Hamiltonian
\begin{align}
    \begin{split}
    \braket{H} &= \frac{\hbar^2 q^2}{m} - \frac{2a\hbar^2}{m} \sum_k \vec{k} \cdot \vec{q} |f_k|^2 \biggl(1 + \prod_{i=1}^3 \biggl( \frac{\sin (\frac{ak_iL}{2})}{ak_i L/2} \biggr)^2 \biggr) \\
    &+ 2\hbar \omega \sum_k |f_k|^2 \biggl( 1 + a^2l^2 k^2 + \prod_{i=1}^3 \biggl( \frac{\sin (\frac{ak_iL}{2})}{ak_i L/2} \biggr)^2 \biggr) \\
    &+ \frac{a^2\hbar^2}{m} \biggl[ \biggl(\sum_k \vec{k} |f_k|^2 \biggr)^2 \\
    &+ 2\sum_{kk'} \vec{k} \cdot \vec{k}' |f_k|^2 |f_{k'}|^2 \prod_{i=1}^3 \biggl( \frac{\sin (\frac{ak_iL}{2})}{ak_i L/2} \biggr)^2 \\
    &+ \frac 12 \sum_{kk'} \vec{k} \cdot \vec{k}' |f_k|^2 |f_{k'}|^2 \biggl(\prod_{i=1}^3 \biggl( \frac{\sin (\frac{a(k_i + k'_i) L}{2})}{a (k_i + k'_i) L/2} \biggr)^2 \\
    &+ \prod_{i=1}^3 \biggl( \frac{\sin (\frac{a(k_i - k'_i) L}{2})}{a (k_i - k'_i) L/2} \biggr)^2 \biggr) \biggr] \\
    &+ 2\sum_k (V_k f_k + V_k^* f_k^*) \prod_{i=1}^3 \frac{\sin (\frac{(1-a) k_i L}{2})}{(1-a)k_i L/2}
    \end{split}
\end{align}
In this formulation there is no well-defined notion of any separation distance between the electrons. As such the only length scale of the system is the one with which we defined our wave function, i.e. the system length $L$. Hence the Coulomb contribution will go as $\braket{1/r} \sim 1/L$ and vanish in the continuum limit. 

Away from $a \approx 0$, the $\sin x/x$ terms will vanish due to phase cancellation. Taking the limit as $a\to 1$,
\begin{align}
    \begin{split} \label{eqn:H_a=1_nakano}
        \braket{H} &= \frac{\hbar^2 q^2}{m} - \frac{2\hbar^2}{m} \sum_k \vec{k} \cdot \vec{q} |f_k|^2 + \frac{\hbar^2}{m} \biggl(\sum_k \vec{k} |f_k|^2 \biggr)^2 \\
        &+ 2\sum_k (V_k f_k + V_k^* f_k^*) + 2\hbar \omega \sum_k(1 + a^2l^2 k^2) |f_k|^2 \\
        &= 2\braket{H}_{LLP}
    \end{split}
\end{align}
The $U$ transformation yields twice the LLP Hamiltonian for a plane wave product state and a corresponding ground state energy of $E_{GS} = -2\alpha \hbar \omega$, following an identical procedure as described in Section \ref{sec:methods} and in \cite{LLP}, verifying that it is the appropriate generalization of the LLP transform.

\section{Comparison with Devreese}
Whereas our wave function (equation \ref{eqn:wfn}) is a product state written in $(\vec{r}_1, \vec{r}_2)$ coordinates, the strong-coupling oscillator-oscillator wave function proposed by Devreese \cite{devreese92} (DEV92) is written as a translationally invariant product of Gaussians dependent only on the relative ($\vec{r} = \vec{r}_1 - \vec{r}_2$) and center-of-mass (COM) ($\vec{R} = (\vec{r}_1 + \vec{r}_2)/2$) coordinates:
\begin{align}
    \begin{split}
        \psi_D(\vec{r},\vec{R}) &= \psi(\vec{r}) \psi(\vec{R}) \\
        \psi_D(\vec{r}) &= \left(\frac{\Omega}{2\pi} \right)^{3/4} e^{-\Omega r^2/4} \\
        \psi_D(\vec{R}) &= \left(\frac{2\Omega_1}{\pi} \right)^{3/4} e^{-\Omega_1 R^2}
    \end{split}
\end{align}
Our proposed wave function (equation \ref{eqn:wfn}) can be written in the same coordinate system:
\begin{align}
    \begin{split}
        \psi(\Vec{r}) &= \frac{1}{\sqrt{2} (2\pi \sigma^2)^{3/4}(1+e^{-y^2/2})^{1/2}} \bigl( e^{-(\frac{\Vec{r}-\Vec{d}}{2\sigma})^2} + e^{-(\frac{\Vec{r}+\Vec{d}}{2\sigma})^2} \bigr) \\
        \psi(\Vec{R}) &= \left( \frac{2}{\pi \sigma^2} \right)^{3/4} e^{-(\frac{\Vec{R} -\Vec{M}}{\sigma})^2}
    \end{split}
\end{align}
where $\vec{M} = (\vec{\mu}_1 + \vec{\mu}_2)/2$ and $\vec{d} = \vec{\mu}_1 - \vec{\mu}_2$. It is straightforward to see that the COM wave function $\psi(R)$ is identical to that of DEV92 when $\Omega_1 = \Omega$, setting $\sigma^{-2} = \Omega$ and $\vec{M}=0$. Similarly, we can obtain $\psi_D(\vec{r})$ from $\psi(\vec{r})$ by fixing $y = d/\sigma = 0$. Defining $\sigma_r$ and $\sigma_R$ as the (independently varying) electron sizes of our relative and COM wave function, respectively, and letting $\sigma_{r,R}^{-2} = \Omega, \Omega_1$, we recover $\psi_D$ identically. 

If we were to naively attempt to introduce an analogous degree of freedom in our wave function by distinguishing $\sigma_r$ and $\sigma_R$, our wave function only be properly symmetrized in the case $\sigma_r = \sigma_R$. This indicates that ideally, we would want to use a more elaborate superposition with different $\sigma$ variables in $(\vec{r}_1, \vec{r}_2)$ space as our wave function. However, in the authors' experience, adding an additional variational parameter causes optimization (using \texttt{scipy.optimize}) to fail; this issue could potentially be resolved with optimization schemes that do not rely on gradient descent, for instance \texttt{scikit-optimize}.

\section{Calculation at arbitrary coupling}\label{app:nakano}
Application of the wavefunction in equation \ref{eqn:wfn} to the Fr{\"o}hlich Hamiltonian given in equation \ref{eqn:ham}, we obtain
\begin{align}
    \begin{split}
    \braket{H} &= \frac{\hbar^2}{2m \sigma^2} \biggl[3- \Delta K \biggr] \\
    &+ 2\hbar \omega \sum_k \biggl(1 + \frac{e^{-a^2k^2\sigma^2/2}}{1 + e^{-y^2/2}} \left( \cos(a\Vec{k} \cdot \Vec{d}) + e^{-y^2/2} \right) \\
    &+ \frac{a^2 \hbar^2 k^2}{2m\omega} \biggr) |f_k|^2 + \frac{2}{1 + e^{-y^2/2}} \sum_k (V_k f_k + V_k^* f_k^*) \\
    &\biggl[ e^{-(1-a)^2 k^2 \sigma^2/4} \biggl( \cos \left( \frac{(1-a) \Vec{k} \cdot \Vec{d}}{2} \right) + e^{-y^2/2} \biggr) \\
    &+ e^{-(1+a^2) k^2 \sigma^2/4} \biggl( \cos \left( \frac{(1+a) \Vec{k} \cdot \Vec{d}}{2} \right) + e^{-y^2/2} \biggr) \biggr] \\
    &+ \frac{e^2}{\epsilon_\infty} \frac{1}{\sigma} \Delta V,
    \end{split}
\end{align}
where we have set the center of mass of the system $\vec{\mu}_1 + \vec{\mu}_2 = 0$. Minimizing with respect to $f_k^*$ yields
\begin{align}
\begin{split}
    f_k &= -\frac{V^*_k}{\hbar \omega (1 + e^{-y^2/2})}\\
    &*\biggl[ e^{-(1-a)^2 k^2 \sigma^2/4} \biggl( \cos \left( \frac{(1-a) \Vec{k} \cdot \Vec{d}}{2} \right) + e^{-y^2/2}\biggr)\\
    &+ e^{-(1+a^2) k^2 \sigma^2/4} \biggl( \cos \left( \frac{(1+a) \Vec{k} \cdot \Vec{d}}{2} \right) + e^{-y^2/2} \biggr) \biggr] \biggl/ \\
    &\biggl(1 + \frac{e^{-a^2k^2\sigma^2/2}}{1 + e^{-y^2/2}} \left( \cos(a\Vec{k} \cdot \Vec{d}) + e^{-y^2/2} \right) + a^2 l^2 k^2 \biggr)
\end{split}
\end{align}
Plugging back in,
\begin{align}
    \begin{split}\label{eqn:nak_ham}
        \braket{H} &= \frac{\hbar^2}{2m \sigma^2} \biggl[3- \Delta K \biggr] +\frac{e^2}{\epsilon_\infty} \frac{1}{\sigma} \Delta V \\
        &- 2\sum_k \frac{V^*_k}{\hbar \omega (1 + e^{-y^2/2})^2}\\
        &*\biggl[ e^{-(1-a)^2 k^2 \sigma^2/4} \biggl( \cos \left( \frac{(1-a) \Vec{k} \cdot \Vec{d}}{2} \right) + e^{-y^2/2}\biggr)\\
        &+ e^{-(1+a^2) k^2 \sigma^2/4} \biggl( \cos \left( \frac{(1+a) \Vec{k} \cdot \Vec{d}}{2} \right) + e^{-y^2/2} \biggr) \biggr]^2 \biggl/ \\
        &\biggl(1 + \frac{e^{-a^2k^2\sigma^2/2}}{1 + e^{-y^2/2}} \left( \cos(a\Vec{k} \cdot \Vec{d}) + e^{-y^2/2} \right) + a^2 l^2 k^2 \biggr)
    \end{split}
\end{align}
In the $a\to 0$ limit, this expression reduces to the Hamiltonian from our strong-coupling calculation in Section \ref{app:strong}. In the $a\to 1$ limit and taking the electron separation distance $y = d/\sigma \to \infty$, we expect our energy to be equivalent to that of two independent, single polarons. To see this, we may rewrite the electron-phonon contribution above in dimensionless units $\ts = \sigma/l$, $u=\cos \theta$, and $z = k\sigma$ as
\begin{equation}
    H_{eph} = -\frac{2\alpha}{\pi \ts} \int_0^\infty dz \int_{-1}^1 du \frac{\left[ 1 + e^{-z^2/2} \cos (zyu) \right]^2}{1 + z^2/\ts^2 + e^{-z^2 /2} \cos (zyu)}
\end{equation}
The denominator of this double integral decays as the Lorentzian $\sim \frac{1}{1+z^2/\ts^2}$ as $z\to \infty$ and as such the integrand above dies out much more slowly at large momenta than in our Gaussian (strong-coupling) calculation. This motivates splitting up the integral into small- and large-$z$ pieces, as
\begin{align}
    \begin{split}\label{eqn:E_nak_num}
        \frac{H_{eph}}{\hbar \omega} &\approx -\frac{2\alpha }{\pi \ts} \biggl[ \int_{z_c}^\infty dz \int_{-1}^1 du \frac{1}{1 + z^2/\ts^2} \biggr] \\
        &+ \int_0^{z_c} dz \int_{-1}^1 du \frac{\left[ 1 + e^{-z^2 /2} \cos (zyu) \right]^2}{1 + z^2/\ts^2 + e^{-z^2 /2} \cos (zyu)} \\
        &\approx -2\alpha \biggl(1 - \frac{2\arctan \left( \frac{z_c}{\ts} \right)}{\pi} \biggr) \\
        &-\frac{2\alpha }{\pi \ts} \int_0^{z_c} dz \int_{-1}^1 du \frac{\left[ 1 + e^{-z^2 /2} \cos (zyu) \right]^2}{1 + z^2/\ts^2 + e^{-z^2 /2} \cos (zyu)}
    \end{split}
\end{align}
where the choice of cutoff $z_c$ depends on where the oscillatory Gaussian factor $e^{-z^2/2} \cos (zyu)$ dies out. Though the angular integral can be evaluated analytically, this rewriting shows clearly the appearance of the $-2\alpha$ energy lowering we expect in the weak-coupling limit; it may be shown numerically that the $\arctan$ term cancels out the double integral in the expression above.

\begin{figure}
    \centering
    \includegraphics[width=0.9\linewidth]{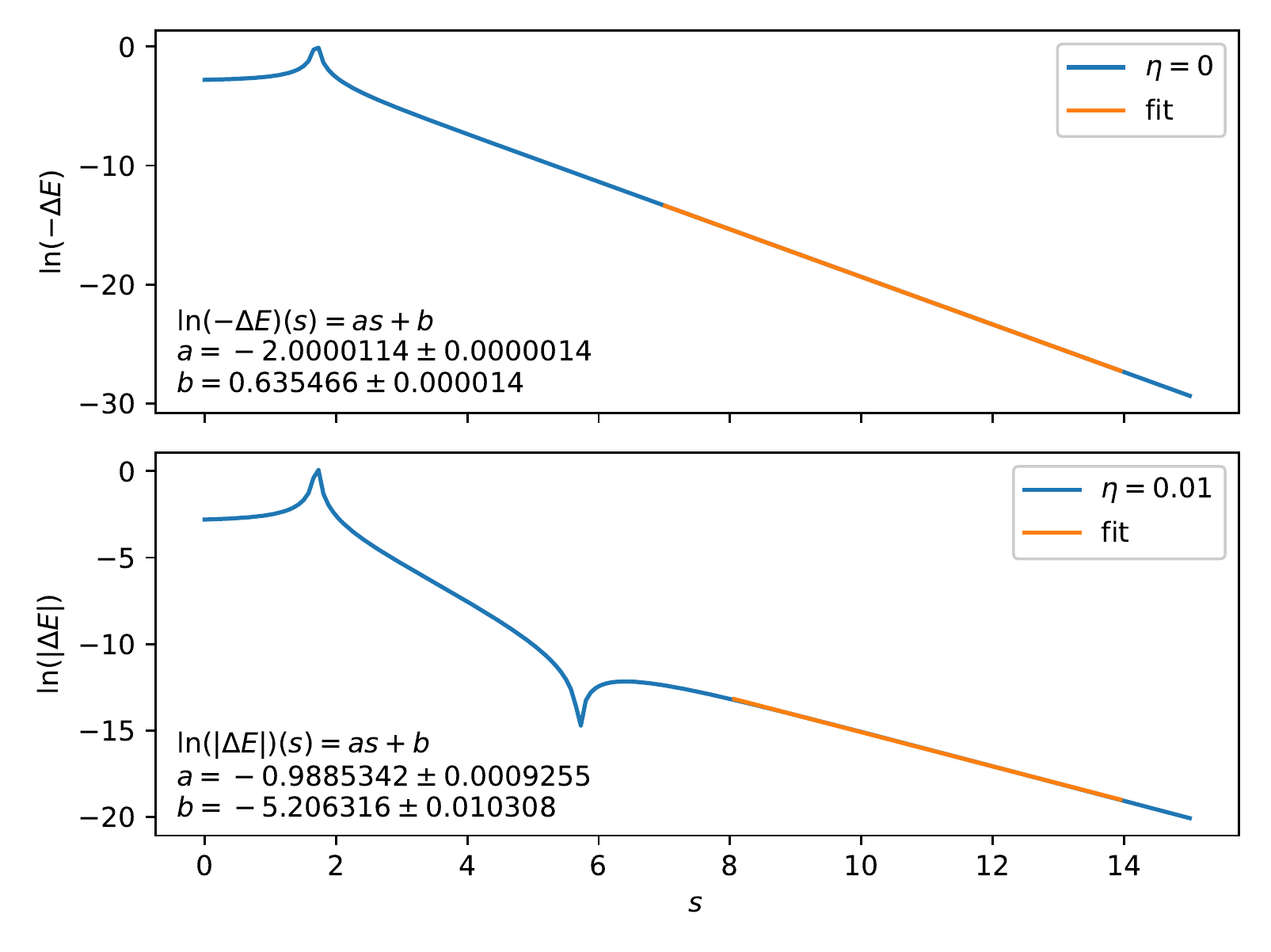}
    \caption{Log-log plots of $\Delta E(\sigma) = (E_{opt} - E_\infty)/|E_\infty|$ for $U=0.1$, a) $\eta=0$ and b) $\eta > 0$. The first cusp in both plots is real and denotes optimization of the binding energy by finite $\ts \sim 1-10$. The second cusp marks the transition from negative to positive binding energies at finite $\eta$.}
  \label{fig:loglog_dE}
\end{figure}

\section{Weak-coupling phenomenology}
\label{app:beta}
\begin{figure}[ht]
  \centering
  \includegraphics[width=\linewidth]{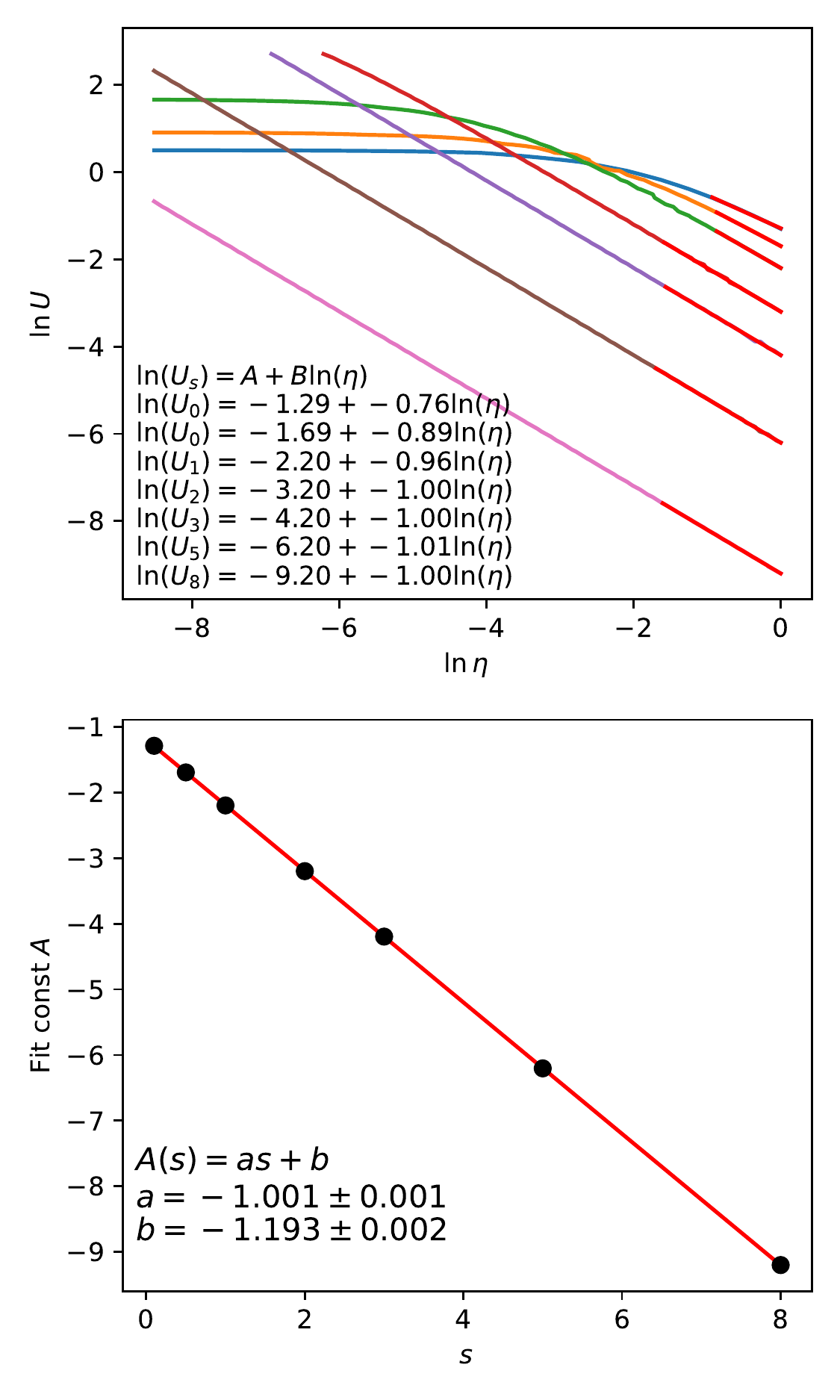}
  \label{fig:sfit_y10}
  \caption{Scaling of fixed-$\sigma$ contours on a (natural log-natural log) $U$ vs $\eta$ phase diagram in weak coupling, with $y_\infty = 10$. Power law behavior gives $U\eta \sim 1/\ts$, in agreement with Fig. \ref{fig:loglog_dE}; see text.}
  \label{fig:wkPD_fits}
\end{figure}

In Fig. \ref{fig:loglog_dE} we plot the normalized binding energy $\Delta E = (E(y=1)-E_\infty)/E_\infty$ as a function of fixed $\ts$. The top panel shows the $\eta = 0$ case of perfect screening. Beyond $\log(\ts)\sim 2$, $\Delta E \sim -\Delta K/\sigma^2$ vanishes in the semiclassical limit (see equation \ref{eqn:nak_ham}). For a particular nonzero $\eta =0.01$, $U = 0.1$,  Fig. \ref{fig:wkPD_fits} shows a window of binding at small $\ts$ with a small regime again approximately controlled by $-\Delta K/\sigma^2$; at larger $\ts$, the repulsive screened Coulomb interaction dominates and $\Delta E \sim 1/\sigma$.

\begin{figure}
    \centering
    \includegraphics[width=\linewidth]{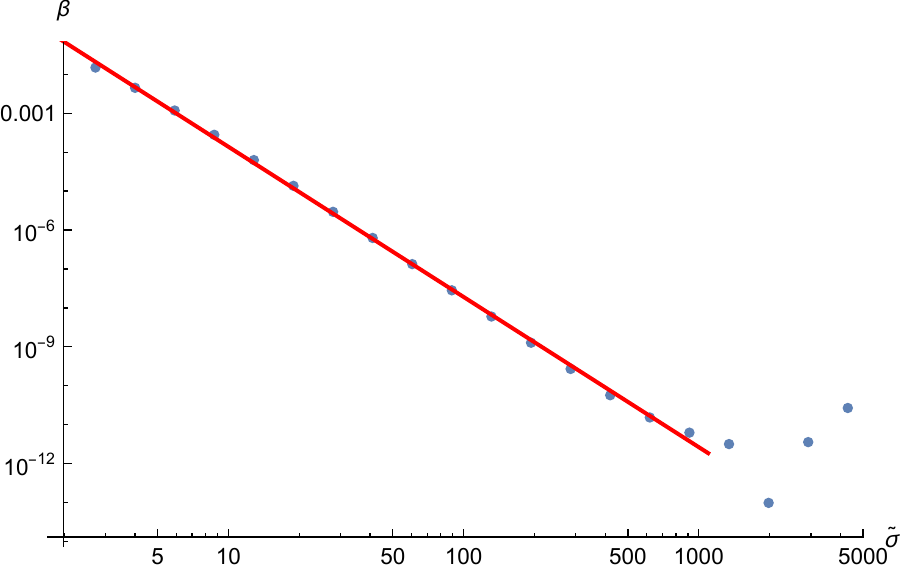}
    \caption{Phonon underscreening parameter $\beta$ away from the semiclassical $\ts-\to\infty$ limit. The fit suggests a power law with exponent $-3.86$.}
    \label{fig:beta}
\end{figure}

For the case of $\tilde\sigma$ extrinsically restricted in the weak-coupling regime, the energy difference between the bipolaron with separation $y$ and two well-separated polarons can be written
\begin{equation}
    \Delta E = -\frac{1}{\ts}\Delta K +\frac{U}{\ts}\Delta V -2\alpha \Delta H_{\text{eph}}.
    \label{eqn:dEgeneral}
\end{equation}
In the weak-coupling regime $a$ will always optimize to $1$, and the electron-phonon energy lowering simplifies to
\begin{equation}
    \Delta H_{\text{eph}} = \Delta\left[ \frac{1}{\pi\ts}\int_0^{z_c}\!dz\,\int_{-1}^1\!du\, \frac{I_0^2}{I_0 +z^2/\ts^2}\right]
    \label{eqn:deltaHeph}
\end{equation}

where
\begin{equation}
    I_0 = 1 +\frac{e^{-z^2/2}(\cos(z y u) +e^{-y^2/2})}{1 +e^{-y^2/2}}
\end{equation}
In the $\ts \to \infty$ limit where the integrand simplifies to $I_0$ it can be shown analytically that $\Delta H_\text{eph} = \frac{1}{\ts}\Delta V$. This motivates rewriting (\ref{eqn:dEgeneral}) as (\ref{eqn:dE}). We have computed $\beta$ by numerical evaluation of $H_\text{eph}(y=1)$ and using $H_\text{eph}(y=\infty) = -2\alpha$. The results are shown in Fig. \ref{fig:beta} on a logarithmic scale, fitted to an apparent power law with exponent $-3.86$ until the results become limited by integration error. 

In Fig. \ref{fig:wkPD_fits}, we obtain a fully numerical phase diagram of the binding energy $\Delta E = (E(y=1) -E(y=500))/|E(y=500)|$ and fit the large-$\eta$ tails of our obtained $\Delta E = 0$ phase boundaries at different $\ts$ to
\begin{equation}
    \log U = A + B\log \eta.
\end{equation}
We reliably find the expected $1/\eta$ behavior. To examine the scaling with $\ts$ at large $\eta$, we plot the constant fit coefficient $A$ as a function of $\ts$ in Fig. \ref{fig:wkPD_fits}, and verify a $1/\ts$ power law in the regime numerically accessible to us.

\bibliography{biblio}

\end{document}